\documentclass[11pt,a4paper]{article}
\usepackage[utf8]{inputenc}
\usepackage{authblk}
\usepackage{color}
\usepackage{hyperref}
\hypersetup{
    colorlinks=true,
    linkcolor=blue,
    urlcolor=blue,
}
\usepackage[sorting=none]{biblatex}
\usepackage[a4paper,top=2.5cm]{geometry}

\title{Software Challenges For HL-LHC Data Analysis}

\author[1]{The ROOT Team\thanks{\href{mailto:rootdev@cern.ch}{rootdev@cern.ch}}: Kim Albertsson Brann}
\author[2]{Guilherme Amadio}
\author[2, 3]{Sitong An}
\author[2]{Bertrand Bellenot}
\author[2]{Jakob Blomer}
\author[4]{Philippe Canal}
\author[2]{Olivier Couet}
\author[2]{Massimiliano Galli}
\author[2]{Enrico Guiraud}
\author[2]{Stephan Hageboeck}
\author[5]{Sergey Linev}
\author[2]{Pere Mato Vila}
\author[2]{Lorenzo Moneta}
\author[6]{Alja Mrak Tadel}
\author[2]{Axel Naumann}
\author[2]{Vincenzo Eduardo Padulano}
\author[2]{Fons Rademakers}
\author[7]{Oksana Shadura}
\author[6]{Matevz Tadel}
\author[2]{Enric Tejedor Saavedra}
\author[2]{Xavier Valls Pla}
\author[8]{Vassil Vassilev}
\author[2,9]{Stefan Wunsch,}

\affil[1]{Luleå University of Technology, Sweden}
\affil[2]{CERN, Geneva, Switzerland}
\affil[3]{Carnegie-Mellon University, USA}
\affil[4]{Fermi National Accelerator Laboratory, Batavia, USA}
\affil[5]{GSI, Darmstadt, Germany}
\affil[6]{University of California San Diego, USA}
\affil[7]{University of Nebraska Lincoln, USA}
\affil[8]{Princeton University, USA}
\affil[9]{Karlsruhe Institute of Technology, Germany}
\date{April 2020}

\begin{document}

\maketitle

\begin{abstract}
    The high energy physics community is discussing where investment is needed to prepare software for the HL-LHC and its unprecedented challenges. The ROOT project is one of the central software players in high energy physics since decades. From its experience and expectations, the ROOT team has distilled a comprehensive set of areas that should see research and development in the context of data analysis software, for making best use of HL-LHC's physics potential. This work shows what these areas could be, why the ROOT team believes investing in them is needed, which gains are expected, and where related work is ongoing. It can serve as an indication for future cooperation and research proposals.
\end{abstract}

\section{Introduction}

HL-LHC's high-precision studies of standard model phenomena and BSM searches will require processing of huge data samples and comparing them to theoretical models with an explosion of parameters. Reducing systematic uncertainties (such as those introduced through correlations) to a level matching the much reduced statistical uncertainties of HL-LHC data requires more accurate and CPU-intensive simulations, data-driven estimations, and testing high-dimensional models, built using a large number of input features and parameters. We expect thus a "superscalar" demand in analysis throughput and data storage that extends well beyond the increase due to higher data statistics.

This document explains why investment in common software is paramount to guarantee the best return for investment by enabling synergy and by keeping duplication of some parts at a level that encourages competition, to counteract segregation of the physics community, to ensure sustainability, and to leverage existing expertise and experience. Investing will enable these projects to serve as hubs of innovation - innovation that is required to address the computing needs of HL-LHC. In this document, the ROOT project explains the benefits of investing in research and development in the areas of data deserialization, data storage size, efficient and transparent use of accelerators and HPC resources, efficient "bridges" between HEP's foundational libraries and relevant external tools, ROOT's interpreter cling as a pivotal foundation, and the graphics system to establish new visual tools for the increased complexity of HL-LHC's physics output.

\section{Significance of ROOT's Input}
ROOT \cite{antcheva2009root} is a fundamental ingredient of virtually all HEP workflows, in areas such as data persistency, modeling, graphics, and analysis. The project demonstrates its community role by its high number of contributors (more than 100 in 2019); more than 1 Exabyte of physics data stored in ROOT files; excellent, active connections with the experiments including direct investment by the experiments; and active exchange with physicists such as more than 50 support messages per average work day in 2019.

ROOT is an open source project following software development's best practices. All contributions are public, as a prerequisite for publicly documenting and recognizing contributions. This openness makes ROOT attractive for funding agencies, as demonstrated by its many contributors, and allows ROOT to serve as a core component of an ecosystem where demonstrators and prototypes for R\&D can "plug in". The project's expertise, its tradition of innovation, and its excellent connections to stakeholders allow the project to establish where investment is needed to harvest the physics potential of the HL-LHC.

\section{Analysis Bottlenecks}
Based on the ROOT team's experience and expertise, and based on discussions with physicists and experiment representatives, the ROOT project predicts the following main challenges that are reflected in this short input document.

\begin{itemize}
    \item{Reading data.
The final steps of analyses are generally limited by the rate at which events are read. This has an effect on computing efficiency as well as physicists' efficiency (time-to-response).}
\item{Data size. Storage needs of data samples and available simulation capacities will limit the availability of simulation samples and as a consequence possibly the quality of physics extracted from HL-LHC's data.}
\item{Efficient use of available compute silicon. Two main aspects will limit the efficiency: lack of accessible programming models for transparent use of heterogeneous and distributed compute resources (accelerators, HPC, but also super-fast, near-CPU memory); use of slow, interpreted languages in performance-critical code at least in analyses.}
\item{Significant use of tools (also from industry) that are inefficient for HEP, for instance for data deserialization, machine learning, modeling, and graphics.}
\item{Visualizations to communicate complex models, correlations, and uncertainties.}
\end{itemize}

\section{Reading Data}

\subsection{Data Size}
The ROOT project believes that the physics potential of the HL-LHC data can only be exploited by disproportionally increasing the size of simulation samples. This is caused to first order by the high demand for sampling high-dimensional parameter spaces of more elaborate models~\cite{madminer}. Or, turning this around: the high statistical power of HL-LHC data enables exclusions of high-dimensional parameter spaces of complex models, which in turn require a higher ratio of simulation over real data than at the LHC. This entails higher demands for analysis and storage, where even today's disk storage demands come at a considerable cost for WLCG, estimated at CHF~50M/year.

While ROOT files have been compared many times over the past decades against possible alternative formats, their applicability and performance characteristics for HEP data remain unrivaled~\cite{blomer2018quantitative} until this day. Although ROOT files (and specifically ROOT’s columnar format, TTree) outperform their competitors for HEP workflows, the ROOT team has identified several improvements so significant that they warrant an evolution of the file format~\cite{blomer2020rootio}. This research and development effort led to a prototype labeled RNTuple. It incorporates many of the successful design decisions of TTree, such as its columnar data layout or horizontal column expansion ("friend trees").

Lossy compression is currently carried out by each experiment separately, tweaking each stored value to match its expected precision in an ad-hoc effort. The community should invest in a sustainable, general and (at least semi-) automatic approach that is central to the common I/O subsystem. A notable research in this area, in collaboration with the ROOT team, is Accelogic's Compressive Computing~\cite{lauret2020compression}.

File format changes (RNTuple) together with improvements in lossless and lossy compression should enable general space savings of 25\%, for all experiments and data formats, without any cost to the quality of the physics results.

\subsection{Read Throughput}
Efficient analyses are dominated by read throughput of the input data. The need for high-throughput reads increases with the use of accelerators and highly parallel analysis workflows.

The ROOT project has determined that two main challenges must be addressed: automatic optimizations of parameters for data serialization and deserialization to not rely on physicists knowing the optimal software configuration, and bulk data processing to increase the amount of data per deserialization instruction count. The latter, together with simple event data models as favored by RNTuple, enable high-throughput data transfer also to GPUs ("structs of arrays" without or with minimal host CPU manipulation) and match data transfer patterns commonly available in High Performance Computing environments.

Other optimizations related to read throughput can have a considerable impact, too. Examples include caching of intermediate analysis results~\cite{watts2015linq} and optimization of the data format / layout to facilities' storage systems (key value stores, distributed multi-node I/O, high-latency remote I/O such as through xrootd), as well as data placement and efficient handling of meta-data (derivation history, calibration constants, luminosity information, or quickly locating a given event). Being able to save such meta-data with, but not necessarily in a file, would enable experiments to easily update such metadata, and make bookkeeping easier.

Reduced file sizes also mean higher deserialization throughput in "events per second". With benefits from caching of intermediary results and optimized data paths (bulk data processing, optimizations for tomorrow's storage systems) we predict possible throughput increases of a factor 2-5 compared to the LHC data throughput, depending on the storage system and analysis workflow.

\section{Efficient Use of Hardware}
Today more than ever, physicists expect to be able to focus on the physics analysis, rather than its coding. To some extent this was triggered by the recent evolution of the Python scientific ecosystem demonstrating that complex analyses can be written in a compact style, using for instance efficient high level Python packages such as NumPy.

Similar patterns can be used in other languages. C++'s traditional verbosity can be alleviated by deferring type information to runtime - thanks to ROOT's C++ just-in-time compiler cling. This allowed ROOT to create a declarative analysis interface, RDataFrame, for writing compact yet efficient analyses in either C++ or Python, exposing the "what" to physicists while hiding the "how" in its implementation details.

\subsection{PyROOT}
We encourage the approach where slow, interpreted languages such as Python are used to compose nonetheless efficient analyses from calls to optimized libraries such as RDataFrame, rather than having the complete analysis written in a slow, interpreted language. Real-world, production analysis mini-frameworks not following this guideline are regularly observed to be 100 times slower than standard workflows\footnote{See for instance \href{https://indi.to/gQL7P}{https://indi.to/gQL7P}}.

We are convinced that investing in ROOT's unique, extremely powerful (automatic) Python bindings can greatly facilitate and accelerate Python analyses. It is an enabling component to create a larger ecosystem with Python and C++ elements. It makes performant C++ code more accessible thanks to simple Python interfaces, allowing more users to rely on these high-performance libraries. This entails defining abstractions that shield the performance-critical parts. Given such abstractions, the use of accelerators is a natural extension.

\subsection{Data Layout}
For optimal throughput and efficiency, a data layout has to take into account the hardware's requirements. It should be implemented behind the scene of a simpler analysis interface such as RDataFrame, where the engine carrying out the analysis steps knows how to optimally schedule and layout data and transfers. The determination of what is "optimal" can happen at runtime, based on the available hardware and on characteristics of the analysis. This is a significant R\&D task, with equally significant potential performance improvements.

\subsection{Domain-Specific Languages}
Even though Python is the language of choice for many analyses, its performance (or lack thereof) and its verboseness when dealing with nested iterations poses a challenge. Domain specific languages (DSLs) promise to solve this to some extent by providing a more compact way of coding. One of the major concerns with DSLs is the inability to debug that language - in general, any DSL invented and exclusively adopted by HEP cannot benefit from an existing tooling market. Nonetheless, ROOT's past use of DSLs (such as those of TTree::Draw and TFormula) proves that DSLs can be successful with limited scope such as for cuts.

An alternative exists: for instance RDataFrame and RVec (a vector-manipulation and computation library), being high-level interfaces, introduce their own concise expression "language" for analysis steps, while still staying in a well-known computing language with its tooling and training ecosystem. We are convinced that simple Python interfaces together with performant C++ libraries and just-in-time compilation are superior alternatives to large-scale use of DSLs.

\subsection{Just-In-Time Compilation}
An integral component of the community's software is ROOT's interpreter cling. It quickly converts C++ code to an executable, in-memory, binary program. Among other roles, it provides information needed to store data structures; it is a prerequisite for Python bindings as well as ROOT's C++ and CUDA interpretation; it enables web-based GUI interaction. ROOT's interpreter also allows efficient evaluation of DSLs by transforming them into C++, a mechanism currently used by ROOT's TFormula. It makes simpler analysis interfaces with runtime type determination possible - crucial for writing simple yet highly efficient analyses.

We are convinced that the community should invest in the cling's just-in-time compilation to further unlock its enormous potential, for instance improving the interaction between python and performance C++ libraries to facilitate their use in analyses, and optimizing code at runtime based on available hardware (for instance through cling's CUDA backend). As cling is at the backbone of the community's data serialization, it is paramount to guarantee maintenance.

\subsection{Categorization and Systematics}
Analysis jobs are typically run numerous times, for testing and bug fixes, to obtain the results, and for evaluating uncertainties and correlations. This wastes computing resources and the time of physicists, because it is often easier to rerun everything than to write an efficient implementation that only computes quantities that changed, or that pools common computations in a computation graph.

To make matters worse, analysis frameworks have mushroomed to help with handling categories and computing uncertainties and correlations. We believe that this can be provided centrally to benefit from common investment. Such tools can increase CPU efficiency by optimizing data flow; by reducing processing runs over all input data; or by caching relevant parts of the input data and intermediary results, for re-use in consecutive runs of the analysis.

\subsection{Modeling}
High energy physics analyses use complex statistical models, correlations, and uncertainties, a challenge that not many sciences have taken upon them. RooFit turned out to be the tool of choice. Now, many alternative solutions are on the market, wrapping either industry libraries or re-implementing parts of RooFit from first principles. All of the currently available solutions have a limited featureset; to the best of our knowledge these competing solutions cannot (and are not claiming to be able to) replace RooFit.

Instead of causing community splits by the adoption of limited competing tools, the community should invest in the renovation of its existing tools, to benefit from existing expertise and from shared maintenance synergy. ROOT has recently shown that this is extremely beneficial, with accelerations of common RooFit operations by factors five and beyond~\cite{hageboeck2020roofit}. This is crucial for HL-LHC's complex models used in analyses and combinations.

We believe that these requirements can be addressed by engaging and coordinating with developers of community tools, and by providing much needed sustainability. Developments should cover streamlined model building, offloading of computations to accelerators, and increased throughput by bulk processing of data.

\subsection{Declarative Analysis}
Many related research elements on the topics above have already started~\cite{cervantesparallelization}, usually centered around ROOT's RDataFrame~\cite{piparo2019rdataframe}, the recent declarative analysis interface, which has already been adopted by a large amount of Run 2 LHC analyses. ROOT proposes to invest in RDataFrame, extending it to handle for instance automatic categorization, and derivation of uncertainties, while hiding implementation details and enabling optimizations, data transfer, and scheduling on heterogeneous and distributed compute backends.

\section{ROOT as Ecosystem}
After more than 20 years of community investment, ROOT is providing much of the common key functionality required by analyses and HEP software. Its key parts have been continuously optimized and measured against alternatives; it has been extended to cover functionality that is of general relevance to the community.

Many of these extensions, such as RooFit or TMVA, were initiated by members of the HEP community. The mechanism enabling this is crucial for HEP and its ability to effectively and inclusively share development. Here, ROOT serves as an open, accessible, and extensible core ingredient. Implementing new features is possible with incremental effort, by extending or replacing parts of the functionality provided by ROOT. This in turn brings such R\&D in scope for university groups and their grant requests. Where such R\&D endeavors are successful, ROOT can enable adoption in the community as a catalyzer and distribution mechanism.

\subsection{Where to Adopt External Tools}
We are convinced that adoption of tools external to HEP can have tremendous benefits. Some require a considerable development effort to make them suitable for HEP, for instance to match existing software ingredients, or to satisfy physicists' expectations and traditions. Other parts of the software ecosystem have properties that are very specific to HEP and best addressed by common HEP software ingredients, for reasons of performance, features, or for keeping in-house expertise. Where this is the case, pivotal HEP solutions (such as the cling interpreter or ROOT's data format) should be shared with industry to aid in sustainability and to share development efforts.

For ROOT, recent examples of adopting industry tools include zstd compression, NumPy Python array management, the OpenUI5 GUI library, CuDNN as CUDA machine learning library, and the MIXMAX random number generator. Each of these required effort to provide highly optimized interoperability with existing software. This effort paid off as these tools were perfect matches for HEP's requirements.

\subsection{Fruitful Competition}
Competition is a prerequisite for progress. It is best created by duplicating parts of the existing ecosystem's functionality and competing in that specific area. This enables smooth integration and adoption by the community. It also allows for benchmarking based on technical merits, by comparing existing functionality with a competing implementation.

The ROOT project sees more and more competition taking a different route, without integration into the existing ecosystem but instead based on external tools. This gets a prototype product out quickly, with minimal investment but without consideration of sustainability; it benefits from extra attention through the use of well-known names; it creates the impression of relevance by benefiting from the external tool's relevance; it can use the often disputed\footnote{Industry seems to value physicists as experts on statistical modeling and data analysis, rather than their knowledge of any given tool that is currently perceived as state of the art. Physicists are best employed as data experts, not tooling engineers.} argument that physicists using the external tool will have higher "market value" in industry. Adoption of these prototypes creates isolated islands of competing technical solutions with limited sustainability and relevance for the community as a whole\footnote{See for instance \href{https://github.com/diana-hep/spark-root}{https://github.com/diana-hep/spark-root} or \href{https://github.com/diana-hep/rootconverter}{https://github.com/diana-hep/rootconverter} that are no longer actively developed.}. Investment becomes scattered, not for the benefit of all; technology expertise gets lost.

HL-LHC is lucky enough to have strong software projects. They can play a coordinating role for contributions. We have demonstrated since decades that this model is beneficial for institutes contributing ("owning" certain software parts), for the projects, and for the community as a whole.

\subsection{Machine Learning}
The community should not develop its own fundamental machine learning tools. It should collaborate with other sciences on improving and growing toolsets. Development efforts should focus on HEP-specific usability and optimization layers for model building and features, such as sculpting, and to interface with HEP's optimized ecosystem, such as ROOT I/O for fast inference.

We believe that the community should embrace TMVA as that bridge between ROOT and external machine learning tools such as scikit-learn, XGBoost, TensorFlow, Keras, mxnet, or PyTorch. The community should invest in TMVA so it can provide customized and targeted interfaces for HEP, with sustainability, performance and ease of use in mind, for instance through production-scale grid-deployable inference of unrivaled performance. ROOT's just-in-time compilation can offer unique features here, hiding much of the complexity of "spelling out" actual models, and optimizing them to available hardware at runtime~\cite{moneta2020tmva, albertsson2020tmva}.

\section{Visualization}
Communication of results is an integral part of any physics analysis, and graphics play a crucial role here. A good visualization engine with good defaults noticeably improves the productivity of physicists. A suitable visual language improves the effectiveness of physics reviews.

\subsection{Web-Based Graphics}
We are convinced that web-based graphics and adoption of external, web-based tools reduce maintenance load, give the community access to a larger pool of potential developers, and make graphics more sustainable and usable (platform independence, local vs. remote). Web-based graphics allow for trivial embedding for instance in online monitoring applications or web-based analysis tools ("notebooks").

\subsection{Competing Solutions}
To our knowledge, no alternative solution offers a comparable feature set. Analyses written in Python are tempted to use the Python packages matplotlib or seaborn. We see the start of a separation of the community, making investment in graphics (better ROOT graphics or better integration of matplotlib or seaborn for HEP's purposes) only relevant to a fraction of the community. While ROOT's graphics system addresses multiple usage patterns (application, online, monitoring, analyses, utmost configurability, publication-ready plots), alternatives are mostly used in python-based analyses. We propose to unify the community again by improving the usability of ROOT's new graphics system to a point where defaults are just right, and the effort of using it (especially from Python) vanishes compared to adjusting alternatives to HEP's needs: doing easy things must be easy, and doing hard things must be readily possible.

\subsection{Visualization of Complex Models}
Visualization becomes even more relevant as the community sets out to refine its "visual language", communicating a high number of uncertainties and correlations in complex models. HL-LHC will require advances in this area; even today's visual communication of results as seen at the LHC has progressed to a complexity that motivates a rethinking of some of our visual language. It is well known that alternative solutions have performance issues with complex graphics\footnote{See e.g. \href{https://matplotlib.org/3.2.1/tutorials/introductory/usage.html\#performance}{https://matplotlib.org/3.2.1/tutorials/introductory/usage.html\#performance} or \href{https://stackoverflow.com/questions/8955869/why-is-plotting-with-matplotlib-so-slow}{why is plotting with Matplotlib so slow?} }.

\section{Baseline Operations}
Healthy software projects must have a long queue of innovation, ranging from R\&D to optimizations. They are generally motivated by sustainability aspects or by insights from providing support or training. Delivering those improves usability and performance significantly: the effect of continuous innovation must not be underestimated. Examples include improved random number generators; optimizations to fight bottlenecks observed in production; simplified installation methods; and adoption of new tools and architectures.

At the same time, support and education are a significant part of any successful software project, essential for being visible, to communicate innovation, and to stay "rooted" in the community for feedback and for understanding the actual needs.

\section{Conclusion}
Agreeing on common software prevented segregation within both the analysis and developer communities. It allowed synergy, preventing needless duplication of efforts, and making investments available to all experiments to maximize their return. It rationalized software engineering, infrastructure, maintenance, and sustainability costs that would have otherwise been spread and repeated, instead of allowing for synergy effects across projects. It enabled incremental R\&D with focused, reasonable effort.

The community's trust and investment in common projects should not come for free: we need competition to measure us against and as an additional source of innovation. But we foresee a schism in the analysis community, centered around the main software players in the field\footnote{See e.g. \href{https://indico.fnal.gov/event/21067/contribution/11/material/slides/0.pdf}{https://indico.fnal.gov/event/21067/contribution/11/material/slides/0.pdf}, slide 39.}. While this started off as "old vs. new," this division now shows multiple facets that favor the spread of uncertainty and misinformation. This carries a cost simply due to the non-technical part of the competition, and causes a fruitless duplication of efforts.

We believe that WLCG / EP-SFT should be strengthened as the community's central hub for common software, computing, and data management solutions. This will enable continued sharing of responsibilities, addressing topics such as sustainability and maintenance, recognition, and community-wide adoption, and very importantly support. This model is extremely successful until this day, with examples such as ROOT's new histogramming being developed at LAL; ROOT's new graphics system being developed at GSI; ROOT's new event display being developed at UCSD; ROOT's lossy compression R\&D taking place at BNL; RooFit being developed at NIKHEF; and ROOT's I/O subsystem being coordinated at Fermilab.

We absolutely encourage competition on a technological basis. Many of ROOT's recent advances came by comparing its performance and usability with that of alternatives. Where appropriate this resulted in adoption of the superior tool (for instance from CINT to clang, from zlib to zstd), or in an implementation that was optimized for interplay with other parts (notably I/O, such as for CuDNN or xgboost) or sustainability (Xtensor, dataframes from the R language).

We believe that ROOT will remain one of the most important software ingredients in HEP for HL-LHC: ROOT's role in the community, its collected and collective expertise, and its ongoing innovation warrants the community's continuous trust. ROOT sees significant challenges for HL-LHC workflows such as analyses. We are convinced that they can be solved best by the community investing decisively in common software.

\newpage

\printbibliography

\end{document}